\documentclass[12pt]{article}

\usepackage{graphicx,epstopdf}

\title{Self-Oscillating\\ Wireless Power Transfer Systems}

\author{Sergei A. Tretyakov$^1$, Constantin R. Simovski$^1$,\\ Constantinos A. Valagiannopoulos$^2$, 
  and Younes Ra'di$^3$}

\date{\small $^1$Aalto University, Finland\\ $^2$Nazarbayev University, Kazakhstan\\ $^3$University of Texas at Austin, USA\\[1em] \today}

\begin{document}

\maketitle

\begin{abstract}
  Conventional wireless power transfer systems consist of a microwave power generator and transmitter located at one place and a microwave power receiver located at a distance. Here we show that wireless power transfer can be realized as a single ``distributed'' microwave generator with an over-the-air feedback, so that the microwave power is generated directly at the place where the energy needs to be delivered. We demonstrate that the use of this paradigm increases efficiency and dramatically reduces sensitivity to misalignments, variations in load and power, and possible presence of obstacles between the source and receiver.    
\end{abstract}

\section{Introduction}

\begin{figure}[h!]
\includegraphics[width=0.8\linewidth]{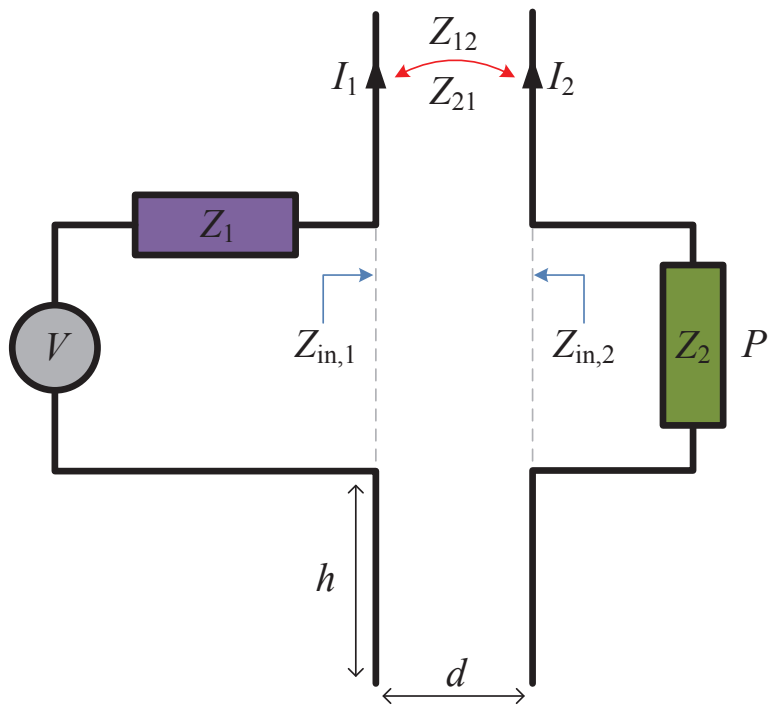}
\caption{A generic scheme of a conventional wireless transfer system.}
\label{fig0}
\end{figure}

In conventional wireless power transfer systems \cite{Science,Bred1,Bred} (developing since the time of Nikola Tesla), the first stage of wireless power transport is a microwave generator which transforms DC or 50/60 Hz power into microwave oscillations. This power transformation is necessary because wireless power links can operate only at reasonably high frequencies. At this stage, any type of microwave generators can be used. Obviously, as in any device which transforms one energy form into another, some power loss is inevitable, although modern microwave generators can be quite efficient. Next, the microwave power available from this generator is send into space using some kind of antenna. Part of this energy is received by another antenna at the receiving end, and the energy is finally converted back to the DC or 50/60 Hz form and used there. Some energy is inevitably lost in the internal resistance of the generator and in the ohmic resistances of the two  antennas. Finally, not all the radiated energy can be captured by the receiving antenna, and some energy escapes into surrounding space. 
This conventional paradigm is illustrated in Fig.~\ref{fig0}. Of course, the two antennas do not have to be electric dipoles: the wire dipoles are used as a generic example of any antenna. 

One of the most important (and unavoidable) reasons for power losses in this system is the dissipation in the internal resistance of the source ($Z_1$). Actually, if the goal is to maximize the power in the load, the source and load impedances should be conjugate matched, and in this case one half of the power is lost in $Z_1$. Moreover, as a matter of fact, in  conventional systems there are \emph{two} such parasitic source resistances, because the microwave generator $V$ is itself fed by a DC or mains source, and there are losses also in the internal resistance of the battery or mains generator. 

In this presentation we will describe an alternative paradigm of wireless power transfer, which completely eliminates parasitic losses in the internal resistance of the microwave power source, because in this new scenario the microwave power is generated directly in the load. One can say that the internal resistance of the microwave generator becomes the load resistance. Moreover, in these new self-oscillating systems, the signal frequency and strength are automatically adjusted due to the generator feedback system, this way minimizing sensitivity to misalignments and variations of the load impedance.

\section{Self-Oscillating Systems}

Let us consider the conventional wireless power transfer system, illustrated in Fig.~\ref{fig0}. Assuming that the wireless link is reciprocal, the mutual impedances are equal and we can denote  $Z_{12}=Z_{21}=-Z_{\rm m}$. The current in the load circuit reads:
\begin{equation}
I_2={Z_{\rm m}V\over (Z_1+Z_{\rm in,1})(Z_2+Z_{\rm in,2})-Z_{\rm m}^2},\label{I}\end{equation}
and  the power delivered to the load is equal to $P=|I_2|^2R_2$, where $R_2={\rm Re}(Z_2)$.

To maximize the amplitude of the current in the load, one brings the circuit to resonance, so that the reactive impedances in the denominator cancel out and the denominator is a real number. In the optimal, idealistic, scenario, the mutual resistance between the two antennas nearly cancels out the radiation resistances [${\rm Re}(Z_{\rm in,1}Z_{\rm in,2}-Z_{\rm m}^2)\rightarrow 0$], which corresponds to the non-radiating mode of two coupled antennas (in the example of two parallel dipole antennas shown in Fig.~\ref{fig0}, this regime corresponds to the mode $I_2=-I_1$ of closely positioned antennas). Note that the difference ${\rm Re}(Z_{\rm in,1}Z_{\rm in,2}-Z_{\rm m}^2)$ is always positive and can only approach zero. Finally, the delivered power can be maximized by matching the load and source resistances, but it is always limited by losses in $Z_1$. 

\begin{figure}[h!]
\includegraphics[width=0.8\linewidth]{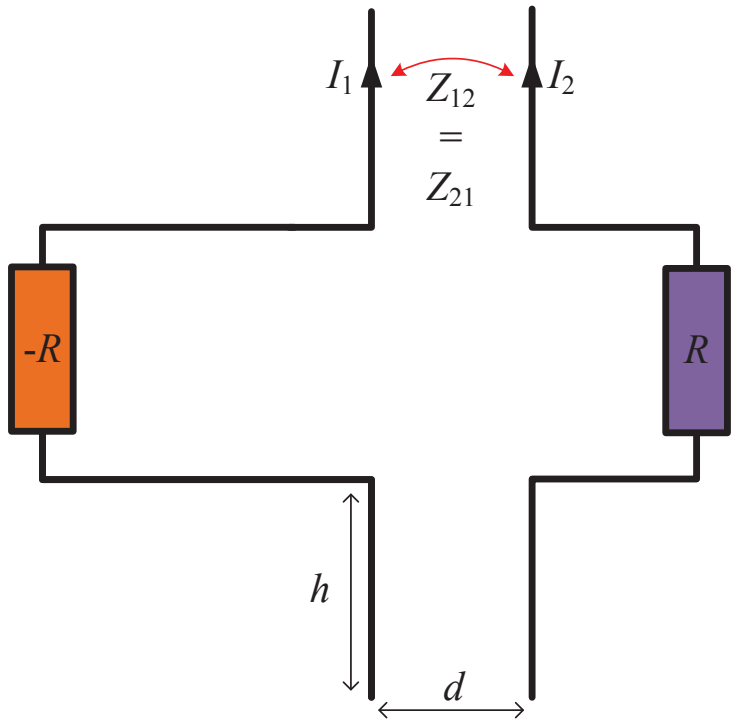}
\caption{A generic scheme of a self-oscillating wireless transfer system.}
\label{fig1}
\end{figure}

Now let us assume that the internal resistance of the source, ${\rm Re}(Z_1)$, can be negative. In this scenario, the absolute value of the denominator in (\ref{I}) can be zero, meaning that in the assumption of linear response, the load current is unbounded. Obviously, this corresponds to a self-oscillating circuit, and  we do not anymore need the voltage source $V$. The energy is delivered directly to the load from a primary power source which creates negative (active) resistance at the active side of the link. This scenario is illustrated in Fig.~\ref{fig1}.

As is well known from the theory of generators based on negative resistance devices, the self-oscillation regime is established ``automatically'', provided that the initial conditions are appropriate. The oscillation level is determined by the non-linearity of the negative-resistance element, while the oscillation frequency corresponds to the resonance of the system. Thus, there is no need for any adjustments of the link if, for instance, the position of the receiver changes. Such change corresponds to changed reactances of the system, which means that if the receiver is moved, the oscillation frequency will change so that the resonance of the system holds, and the delivered power will be optimized automatically. Naturally, receiver positions and load impedance can be changed within some limits, ensuring that the self-oscillations remain possible.

\begin{figure}[h!]
\includegraphics[width=\linewidth]{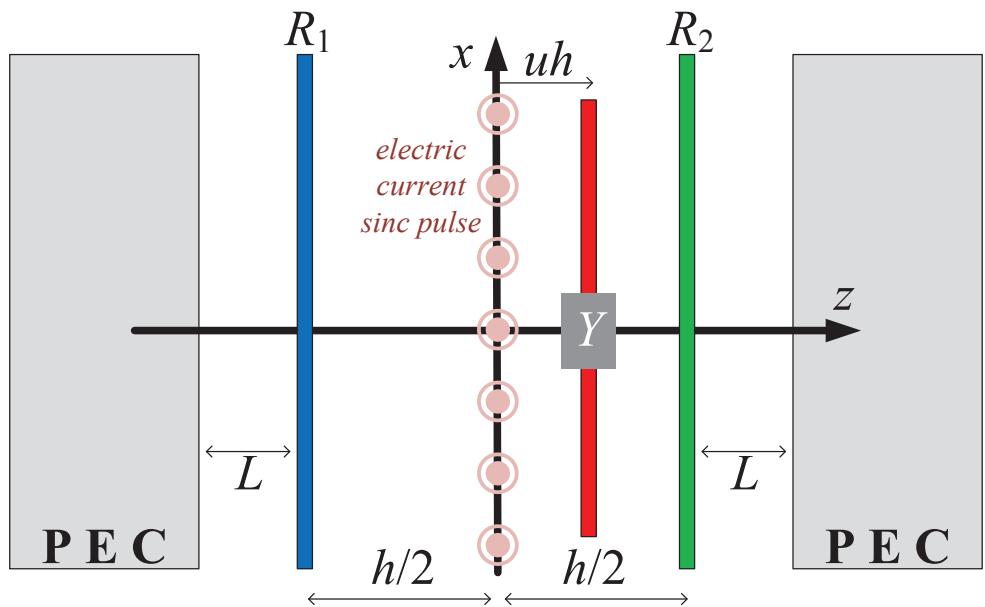}
\caption{One-dimensional model: Admittance $Y$ models an absorbing and scattering obstacle between the source and receiver.}
\label{setup}
\end{figure}

We have modelled the proposed self-oscillating wireless power transfer system using a one-dimensional transmission-line model. The active element is a resistive sheet $R_1<0$, the receiver is a resistive sheet  $R_2>0$, and the regime of compensated radiation into free space is ensured by short-circuit terminations at both ends. The presence of absorbing obstacles between the source and receiver as well as parasitic radiation leakage is modelled by a complex admittance sheet $Y$ positioned at an arbitrary distance between the source and the load. This model set-up is shown in Fig.~\ref{setup}. The self-oscillating regime is initiated by an electric-current sinc pulse at $z=0$.

\begin{figure}[h!]
\includegraphics[width=\linewidth]{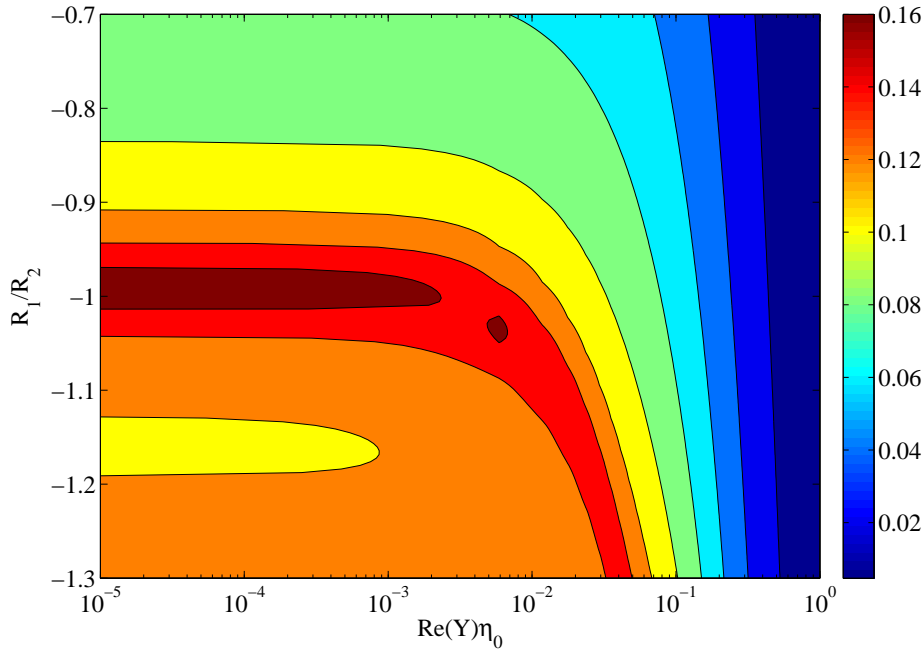}
\caption{One-dimensional model: Robustness  of the self-oscillating wireless power delivery in presence of an absorbing obstacle between the source and receiver.}
\label{results}
\end{figure}

The calculated results are shown in Fig.~\ref{results}. Here, the wireless coupling between the source and the receiver  is due to plane-wave propagation between the active sheet $R_1$  and the receiving resistive sheet $R_2$. The surface admittance  of the obstacle is normalized to the free-space wave impedance $\eta_0$. 
The quantity represented in Fig.~\ref{results}  is the time-averaged electric field (measured in V/m) at the receiver location. The field is averaged over a long period far from $t=0$.   The Fourier transform of the initiating pulse field is a gate function (a low-frequency sinc pulse in time domain). Its amplitude is 1 (measured in V/(m*sec)) for $-\omega_0<\omega<\omega_0$ and 0 elsewhere (indicatively,  $\omega_0/(2\pi)= 3$ GHz). 

Time-averaged electric field strength at the receiver position is shown by color distribution. We see that the delivered power does not change if we introduce an obstacle and increase its conductance up to a certain threshold. Similar property is observed also if we change the position of the obstacle or the receiver, as will be shown in the presentation. These results confirm   robustness  of the proposed system to variations of the environment. 

Note that the proposed system resembles recently conceptualized parity-time (PT) symmetric systems, where gain is compensated by loss in symmetrically positioned active and lossy elements. Although in our proposed devices  ideal PT-symmetry is not required, it is interesting to observe similarities with energy teleportation through nearly opaque screens in PT-symmetric systems \cite{teleport}.

\section{Discussion and conclusion}

Although we have introduced and explained the main idea of self-oscillating wireless power transfer based on the use of negative-resistance  circuits, the same principle can be realized using other  self-oscillating circuits. For example, consider a generator formed by a microwave amplifier with an appropriate positive feed-back loop. In this  alternative scenario the wireless link can be a part of the feed-back circuit of a generator, which creates microwave oscillations directly where the power is needed.   Conceptually, to convert  a conventional generator into a wireless power delivery system, one can let a part of the feed-back signal propagate in space and insert the object to which we want to deliver power into the field of the feed-back electromagnetic wave. 

In conclusion, we have described an alternative paradigm of wireless power transfer, where microwave energy is generated directly at the location where it is needed. The wireless link is a part of the feed-back loop of a microwave self-oscillating circuit. In this scenario, the whole system is a single microwave generator, which directly converts DC or mains power into microwave power \emph{at the position of the receiver}.



\end{document}